\documentclass[aip,preprint,floatfix]{revtex4-1}  

\usepackage[utf8]{inputenc}
\usepackage[T1]{fontenc}
\usepackage{amsmath,amsfonts,amssymb}
\usepackage{graphicx}
\usepackage{epstopdf}
\usepackage{subcaption}
\usepackage{mathtools}
\usepackage{siunitx}
\usepackage{textgreek}
\graphicspath{{figures/}}
\usepackage{xr-hyper}
\usepackage[capitalize]{cleveref}
\usepackage{units}
\usepackage{xcolor}
\usepackage{csquotes}
\MakeOuterQuote{"}

\makeatletter
\newcommand*{\addFileDependency}[1]{
  \typeout{(#1)}
  \@addtofilelist{#1}
  \IfFileExists{#1}{}{\typeout{No file #1.}}
}
\makeatother

\newcommand*{\myexternaldocument}[1]{%
    \externaldocument{#1}%
    \addFileDependency{#1.tex}%
    \addFileDependency{#1.aux}%
}

\myexternaldocument{supp} 

\DeclareUnicodeCharacter{221E}{$\infty$}  

\renewcommand{\vec}[1]{\mathbf{#1}}

\DeclarePairedDelimiter\abs{\lvert}{\rvert}%
\DeclareMathOperator{\tr}{tr}

\makeatletter
\def\@email#1#2{%
 \endgroup
 \patchcmd{\titleblock@produce}
  {\frontmatter@RRAPformat}
  {\frontmatter@RRAPformat{\produce@RRAP{*#1\href{mailto:#2}{#2}}}\frontmatter@RRAPformat}
  {}{}
}%
\makeatother

\begin{document} 
\title[Theoretical comparison of real-time feedback-driven single-particle tracking techniques]{Theoretical comparison of real-time feedback-driven single-particle tracking techniques}
\author{Bertus van Heerden}
\affiliation{Department of Physics, University of Pretoria, 0002 Pretoria, South Africa}
\affiliation{Forestry and Agricultural Biotechnology Institute (FABI), University of Pretoria, 0002 Pretoria, South Africa}
\affiliation{National Institute for Theoretical and Computational Sciences (NITheCS), South Africa}
\author{Tjaart P.J. Krüger}
\email{tjaart.kruger@up.ac.za}
\affiliation{Department of Physics, University of Pretoria, 0002 Pretoria, South Africa}
\affiliation{Forestry and Agricultural Biotechnology Institute (FABI), University of Pretoria, 0002 Pretoria, South Africa}

\begin{abstract}
Real-time feedback-driven single-particle tracking is a technique that uses feedback control 
to enable single-molecule spectroscopy of freely diffusing particles in native or near-native 
environments. A number of different RT-FD-SPT approaches exist, and comparisons between 
methods based on experimental
results are of limited use due to differences in samples and setups. In this study, we used 
statistical calculations and dynamical simulations to directly compare the performance of 
different methods. The methods considered were the orbital method, the Knight`s Tour (grid 
scan) method and MINFLUX, and we considered both fluorescence-based and interferometric 
scattering (iSCAT) approaches. There is a fundamental trade-off between precision and speed, 
with the Knight’s Tour method being able to track the fastest diffusion but with low 
precision, and MINFLUX being the most precise but only tracking slow
diffusion. To compare iSCAT and fluorescence, different biological samples were considered, 
including labeled and intrinsically fluorescent samples. The success of iSCAT as compared to 
fluorescence is strongly dependent on the particle size and the density and photophysical 
properties of the fluorescent particles. Using a wavelength for iSCAT that is negligibly 
absorbed by the tracked particle allows an increased
illumination intensity, which results in iSCAT providing better tracking for most samples. 
This work highlights the fundamental aspects of performance in RT-FD-SPT and should assist 
with the selection of an appropriate method for a particular application. The approach used 
can easily be extended to other RT-FD-SPT methods.
\end{abstract}
\maketitle

\section{Introduction}
\label{sec:intro}  

Single-molecule spectroscopy (SMS) gives access to the properties and dynamics of biomolecules and other molecular or nanoscaled systems that are normally averaged out in bulk experiments. It allows the real-time detection of dynamics that are unsynchronized across separate particles, thus circumventing ensemble averaging and unveiling many hidden details in molecular processes. SMS typically measures the fluorescence emitted by a fluorescent probe or the autofluorescence of a molecule or nanoparticle and allows the spectrum, intensity, lifetime, and polarization of the fluorescence to be detected simultaneously~\cite{gruberIsolatedLightharvestingComplexes2018}. The use of a confocal pinhole, total internal reflection~\cite{toomreLightingCellSurface2001,royPracticalGuideSinglemolecule2008}, or two-photon excitation~\cite{mertzSinglemoleculeDetectionTwophotonexcited1995,zipfelNonlinearMagicMultiphoton2003} allows a smaller observation volume to be probed, 
thus reducing the background noise and enabling the measurement of faint single 
emitters~\cite{nieProbingIndividualMolecules1994}. Example applications include protein 
conformational  dynamics~\cite{mazalSinglemoleculeFRETMethods2019}, transcription in single 
DNA molecules~\cite{kapanidisInitialTranscriptionRNA2006,houRealtime3DSingle2020a}, enzyme
reactions~\cite{haSinglemoleculeFluorescenceSpectroscopy1999, 
jiangSensingCooperativityATP2011}, and light-harvesting complexes switching between different 
functional states~\cite{krugerControlledDisorderPlant2012a,schlau-cohenSingleMoleculeIdentificationQuenched2015,gwizdalaControllingLightHarvesting2016b,krugerHowReducedExcitonic2017}. 

During the past three decades, unprecedented advances have been made in the application of SMS to numerous types of biomolecules but its full potential remains to be developed. This is mainly due to its limitations not having been well addressed yet. One major limitation of standard approaches to 
SMS is its requirement to remove a biomolecule from its natural environment. Performing SMS \textit{in vivo} is a highly challenging endeavor, considering the extremely high packing density of macromolecules in their native environment, which puts a significant constraint on the ultrahigh selectivity and sensitivity required for a successful 
single-molecule experiment. For this reason, SMS is commonly performed on a highly 
purified sample of molecules in an \textit{in vitro} environment. Furthermore, collection of a 
statistically significant number of fluorescence photons requires a sufficiently long 
measurement time. Therefore, to suppress the diffusive motion of biomolecules, they need 
to be immobilized on a substrate, in an enclosing matrix, or in solution, or confined to a small subspace like a liposome or vesicle. Immobilization is commonly done by attaching a highly diluted concentration of molecules to a microscope coverslip through the use of molecular linkers, or by spin-coating the molecules in a polymer host matrix. Molecules can also be trapped in solution using physical, optical, electrokinetic, or thermodynamic principles~\cite{bespalovaSinglemoleculeTrappingMeasurement2019}. Immobilization or confinement of biomolecules allows the use of a tightly focused light source to probe it for extended periods of time. However, this is a highly artificial environment and therefore strongly limits the biological relevance of these experiments. Immobilization of biomolecules may also introduce non-physiological interactions and structural distortions.

One key step closer to \textit{in vivo} SMS is a technique that allows the real-time detection of freely diffusing single molecules for extended periods of time. This can be done by real-time feedback-driven single-particle tracking (RT-FD-SPT), which uses feedback control to keep
a particle in the observation volume, without disturbing the particle's local 
environment~\cite{houRealtime3DSingle2019, 
vanheerdenRealTimeFeedbackDrivenSingleParticle2021}.  RT-FD-SPT is not to be confused with conventional,
image-based single particle tracking, which is also able to measure a particle's motion but 
uses a wide field of view that does not allow for concurrent spectroscopy. RT-FD-SPT has, for
example, been used to measure fluorescence 
lifetimes~\cite{wellsTimeResolved3D2010,chenMeasuringDNAHybridization2019},
spectra~\cite{hellriegelRealtimeMultiparameterSpectroscopy2009a}, and 
antibunching~\cite{wellsTimeResolved3D2010,chenMeasuringDNAHybridization2019} of freely diffusing molecules
in live cells. RT-FD-SPT also has better time resolution and a much longer tracking 
range~\cite{vanheerdenRealTimeFeedbackDrivenSingleParticle2021} than conventional image-based SPT.

Despite a rather slow development of RT-FD-SPT methods during the past ca. two decades compared to other single-molecule techniques~\cite{vanheerdenRealTimeFeedbackDrivenSingleParticle2021}, there are already many different RT-FD-SPT methods in use, and accurately evaluating the
performance of different methods is key to choosing the right method for a specific
application. Comparisons can be made using experimental
results~\cite{houRealtime3DSingle2019, 
vanheerdenRealTimeFeedbackDrivenSingleParticle2021}, but this is not ideal due to 
considerable variation in experimental setups and samples. Some methods have also been investigated theoretically, using either statistical analysis~\cite{shenOptimalMeasurementConstellation2012,gallatinOptimalLaserScan2012,balzarottiNanometerResolutionImaging2017,zhangInformationEfficientOffCenterSampling2021,vickersInformationOptimalControl2021,masulloCommonFrameworkSinglemolecule2022} or dynamical simulations~\cite{berglundFeedbackControllerDesign2004,anderssonTrackingSingleFluorescent2005,berglundPerformanceBoundsSingleparticle2006,anderssonLinearOptimalControl2009a, wangOptimalStrategyTrapping2010,fieldsOptimalTrackingBrownian2012} but this has so far only been done for a limited number of methods and using approaches or criteria that do not allow a direct comparison of all RT-FD-SPT methods. We started to address this shortcoming in a previous 
study~\cite{vanheerdenTheoreticalComparisonRealtime2021b}, wherein a comparison of different RT-FD-SPT methods was made using statistical analysis. In this work, we
expand that study to more thoroughly investigate three commonly used RT-FD-SPT methods and extend the
comparison using a dynamical simulation. We consider both fluorescence emission and scattering as photon sources. Interferometric scattering (iSCAT) is considered for the latter. Due to the limited theoretical development for iSCAT-based tracking methods we first derived an expression for its Cramér-Rao bound before evaluating its performance. Overall, our study provides a direct comparison of
the performance of different RT-FD-SPT methods. It also illustrates some fundamental aspects of RT-FD-SPT performance that are often overlooked in experimental literature or studies purely focused on localization (e.g., Ref.~\onlinecite{masulloCommonFrameworkSinglemolecule2022}), such as an important trade-off in RT-FD-SPT between tracking precision and tracking speed, and the relationship between localization precision and realized tracking error, specifically in the context of Kalman filtering. 

In \Cref{sec:possens} we describe the different position sensing methods that were investigated and 
derive an expression for the iSCAT Cramér-Rao bound.  \Cref{sec:results} shows the results of the
statistical analysis and dynamical simulations, and \Cref{sec:dissconc} provides a discussion of 
the results as well as concluding remarks.

\section{Position Sensing} 
\label{sec:possens}

RT-FD-SPT is based around a feedback system that uses a measured particle position to continually re-center the particle in the observation volume. Such a feedback system consists of a "position sensor", a control system, and an output actuator. Approaches differ mainly in how the particle position is measured~\cite{vanheerdenRealTimeFeedbackDrivenSingleParticle2021}. In this
study, we focus on single-detector methods that can be used in two dimensions. This is done for simplicity and due to the fact that, while iSCAT is capable of 3D tracking when the motion is sufficiently slow~\cite{taylorInterferometricScatteringMicroscopy2019c, sandoghdarConfocalInterferometricScattering2022}, large axial motion would not easily be tracked in an RT-FD-SPT context. The three methods considered are also well-suited to illustrate the fundamental aspects of performance that this study aims to highlight. The
analysis can, however, easily be extended to three dimensions or other position sensing
methods.  The 2D results are immediately applicable to scenarios such as diffusion in flat lipid bilayers.

Three position-sensing methods commonly used in two dimensions are the
orbital~\cite{enderleinTrackingFluorescentMolecules2000}, Knight's Tour~\cite{wangOptimalStrategyTrapping2010}, and MINFLUX~\cite{balzarottiNanometerResolutionImaging2017} methods. The former two involve a beam scanning pattern over a semi-continuous area, while MINFLUX involves a "constellation" of a small number of points. All three methods have been extended to three dimensions~\cite{levi3DParticleTracking2005,
gwoschMINFLUXNanoscopyDelivers2020, houRobustRealtime3D2017}. These methods have thus far
been applied using fluorescence measurements, but we will also consider them in the
context of the technique called interferometric scattering
(iSCAT)~\cite{lindforsDetectionSpectroscopyGold2004a}, which uses scattered light.

\subsection{Scanning patterns}
\begin{figure*}[!htb] 
	\center{\includegraphics[width=0.8\textwidth]{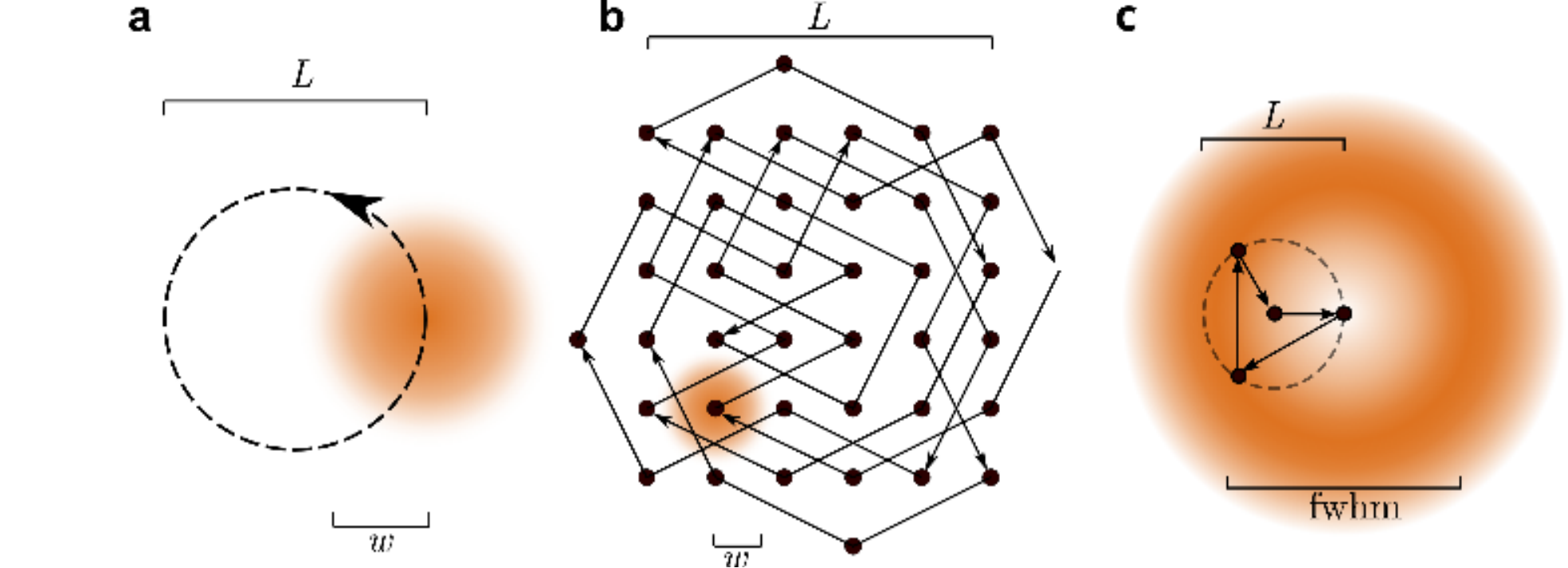}}
	\caption{\label{orb}(a) Orbital scanning pattern with beam waist $w$ and 
	scanning length $L$. (b) Knight's Tour scanning pattern showing the scanning 
	length $L$ and beam waist $w$.  (c) MINFLUX scanning pattern, showing scanning length $L$ and
	beam size parameter $\mathrm{fwhm}$. The particle is kept in the center of the 
	scanning pattern in each case.}
\end{figure*}

In the orbital method, a Gaussian laser spot is scanned in a circle around the particle of interest
(\cref{orb}a). If the particle is not in the center and the orbital frequency is sufficiently high, the detected intensity varies
sinusoidally, enabling one to use the amplitude and phase of this variation to determine the
particle position in the $x$--$y$ plane~\cite{enderleinTrackingFluorescentMolecules2000,
berglundFeedbackControllerDesign2004, kis-petikovaDistanceMeasurementCircular2004}. For
sufficiently large values of the orbital radius $R = L/2$, where $L$ denotes the orbital diameter or "scanning length" (see \cref{orb}), the average detected intensity has a local minimum when the particle is in the center of the orbital. For small $R$, there is a maximum average intensity in the center. A value of $R = w / \sqrt{2} \approx 0.7w$, where $w$ is the beam waist, is the
largest value of $R$ for which there is a maximum average intensity in the center~\cite{enderleinTrackingFluorescentMolecules2000}, and this has been shown to be the optimal geometry to maximize position 
information~\cite{berglundPerformanceBoundsSingleparticle2006,
gallatinOptimalLaserScan2012}.

The Knight's Tour method also uses a Gaussian laser spot, but instead of scanning in a
circle, the spot is scanned across a grid of points by means of a "knight's tour"
(\cref{orb}b). This is a heuristic to cycle through the points of the grid in a way that continually samples the entire area, rather than from one side to the other as in a raster scan. This allows feedback to be applied on every photon detection without the need to
wait for a scan to complete, which would improve the tracking of fast-moving particles that might move across the scan area faster than a raster scan does. The Knight's Tour method was originally developed as a way of increasing the size of the scan area and thus increasing the speed at
which particles can be tracked. Before its implementation, a larger scan area was obtained by increasing the beam waist with the orbital method, which decreases the tracking precision~\cite{berglundPerformanceBoundsSingleparticle2006,gallatinOptimalLaserScan2012}. In this work, we use the knight's tour pattern from the first work by Wang and Moerner~\cite{wangOptimalStrategyTrapping2010}. Different scan patterns have been used~\cite{wangAdaptiveAntibrownianElectrokinetic2011,
fieldsElectrokineticTrappingOne2011},
and it has also been extended to three dimensions by the Welsher group in a method called
3D-DyPLoT~\cite{houRobustRealtime3D2017,
houRealtime3DSingle2019,houRealtime3DSingle2020a}. 

MINFLUX utilizes an illumination minimum -- typically using a doughnut-shaped laser beam spot -- to
measure the particle position (\cref{orb}c). Keeping the particle in the center of the illumination minimum minimizes the total number of detected photons, enabling more position information to be encoded, in principle, in each detected photon~\cite{balzarottiNanometerResolutionImaging2017,
eilersMINFLUXMonitorsRapid2018a}. This concept is in contrast to the other two methods, where the number of detected photons is maximized. The standard two-dimensional MINFLUX approach utilizes four scan points and the original implementation involved fast scanning of the beam. A recently developed
version,  called p-MINFLUX~\cite{masulloPulsedInterleavedMINFLUX2021}, uses interleaved
laser pulses, which eliminates the need for fast beam scanning but, instead, requires
a fast pulsed laser and time-gating electronics.

\subsection{iSCAT}

RT-FD-SPT has traditionally been based on fluorescence, but it is also possible to make
use of scattered light by using the technique called interferometric scattering (iSCAT).
This technique uses the interference of scattered and reflected light, which, for small particles, provides a significantly stronger signal than the pure scattering signal. The
principle is as follows~\cite{lindforsDetectionSpectroscopyGold2004a}: an incident light
field with electric field amplitude $E_i$ on a particle close to an interface (for example the glass-water interface of
a coverslip close to the particle) produces a scattered field with amplitude $E_s$ from the particle as well as a reference field $E_r$ from the reflection at the interface. These two fields
interfere, giving rise to a detected intensity $I_{det}$ given by 
\begin{align}
I_{det} \propto \abs{rE_i + sE_i}^2 &= \abs{E_i}^2 (r^2 + 2 \abs{r}\abs{s}\cos{\theta} +
\abs{s}^2), 
\end{align}
where $r$ is the reflectivity of the interface, $s$ is the complex scattering
coefficient, and $\theta$ is the difference in phase between the reflected and scattered
fields. If the illumination originates from a laser beam with cross-sectional area $A$, $s$ 
can be determined from the scattering cross-section 
$\sigma_{scat}$ through the relation $\abs{s}^2 = \sigma_{scat} / A$.

Due to the square on $\abs{s}$, for a small particle the pure scattering term,
$\abs{E_i}^2\abs{s}^2$, is negligible compared to the interference term, $\abs{E_i}^2\abs{r}
\abs{s} \cos{\theta}$. Even though the detected signal is dominated by the reference term,
$\abs{E_i}^2  r^2$, the latter should remain constant and can be estimated by measuring the background $N_{bkg}$ when there is no particle in the observation volume. The actual iSCAT measurement is then made in terms of the contrast $C_f$ between the
interference and the background, defined as
\begin{align} 
\label{conteq}
C_f = \abs*{\frac{N_{det} -
N_{bkg}}{N_{bkg}}}, 
\end{align}
where $N_{det}$ is the number of photons detected during the measurement.

iSCAT essentially provides only a different source of photons and the actual position
measurement can thus still be done using a variety of methods. Traditionally, the method
has used a
camera~\cite{gemeinhardtLabelfreeImagingSingle2018,
jacobsenInterferometricOpticalDetection2006,lindforsDetectionSpectroscopyGold2004a}
but it has recently also been implemented using a knight's tour scan and real-time
feedback~\cite{squiresInterferometricScatteringEnables2019}. The combination of iSCAT with RT-FD-SPT is a promising but as yet untapped area of development~\cite{
vanheerdenRealTimeFeedbackDrivenSingleParticle2021}.

\subsection{Static localization}

A particle that moves slowly in comparison with the feedback bandwidth can be tracked well enough to be considered approximately static relative to the tracking beam. For such a tracking situation it is useful to determine the accuracy with which a static particle's position can be determined. The main
limitations on the accuracy of any particular method are the number of detected photons and
the signal-to-background ratio (SBR). These two factors are usually strongly related but do not necessarily show a one-to-one relationship. We therefore develop a framework to compare the precision of the different position sensing methods given a certain number of detected photons and a certain SBR. To this end, we use the Fisher information, which provides a measure of the amount of information that an observed random variable carries about an unknown parameter. In our case, the observed variable is the photon
counts and the unknown parameter is the particle position. We are ultimately interested
in the Cramér-Rao bound (CRB), which can be calculated from the Fisher information. The CRB is a lower bound on the variance of any position estimator, and thus a direct measure of precision. For fluorescence-based localization, the CRB calculation is described in
Ref.~\onlinecite{balzarottiNanometerResolutionImaging2017}. 

For iSCAT, no CRB has as yet been determined to the best of our knowledge. The major difference 
between the photon statistics of iSCAT and fluorescence is that for iSCAT, the statistics are determined by the background. This is because the interference counts are not measured directly, but measurements are made in terms of the contrast $C_f$ between the interference and background counts. This contrast measurement is dominated by the reference (background) counts, and the uncertainty in
the estimate of the interference counts is therefore mostly determined by the photon noise of the reference counts. Since the reference term is produced purely by the laser, we can assume that the reference counts $N_r \approx N_{bkg}$ are Poisson distributed. We further assume that $N_r$ is large, and it can thus be well approximated by a Gaussian distribution with
variance $N_r$. This leads to the estimate of the interference counts 
$N_{int} \approx N_{det} - N_{bkg}$
having a variance of $\sigma_N^2 = \sigma_{det}^2 + \sigma_{bkg}^2 \approx 2 
\sigma_{bkg}^2 \approx 2N_r = 2N_{int}/C_f$. Note that this variance is different 
from the Poisson variance $\sigma_{N,Poiss}^2 = N_{int}$ that would be applicable if it was possible to 
measure the interference counts directly. As as $C_f$ is typically less than 5\%,
the variance $\sigma_N^2$ is large compared to the mean. We derive an 
expression for the iSCAT CRB using this variance $\sigma_N^2$.

We follow a modified approach of Ref.~\onlinecite{balzarottiNanometerResolutionImaging2017} and 
consider a particle
at a position $\bar{r}_m \in \mathbb{R}^d$ probed with an illumination pattern comprising $K$ 
distinct intensities
$\{I_0(\bar{r}), ..., I_{K-1}(\bar{r})\}$, with a resulting collection $\bar{n} = \{n_0,
n_1, ..., n_{K-1}\}$ interference photons per pattern point and a total number of
interference photons $ N_{int} = n_0 + n_1 + ... + n_{K-1}. $

Assuming $n_i$ to be sufficiently large, $\bar{n}$ is approximately Gaussian with mean
$\hat{N}_{int}$ and standard deviation $\sigma_N$. We can approximate
$n_i$ as also being Gaussian with expected values $p_i \hat{N}_{int}$ and standard deviations
$p_i \sigma_N$, where 
\begin{equation} 
	\label{parameq}
	p_i = \frac{I_i}{\sum^{K-1}_{j=0} I_j},
\end{equation}
with $0 \leq i \leq K-1$.

Thus the probabilities for measuring $\bar{n}$ conditioned to $\hat{N}_{int}$ are given by
\begin{align}
	P(\bar{n}|\hat{N}_{int}) &= \prod_{i=0}^{K-1} \frac{ \exp{-\frac{1}{2}\left(\frac{n_i -
	\hat{N}_{int}p_i}{\sigma_N p_i} \right)^2}}{\sigma_N p_i \sqrt{2\pi}},
\end{align}
where $p_{K-1} = 1 - \sum^{K-2}_{j=0} p_j$. The Fisher information is therefore given by
\begin{align}
	\{F_{\bar{p}}\}_{ij} &= E \left(-\frac{\partial^2}{\partial p_i \partial
	p_j} \ln{P(\bar{n} | \hat{N}_{int})}\right) \text{with } i, j
	\in [0, ..., K-2]\\
	&= \frac{\hat{N}^2_{int} - \sigma_N^2}{\sigma_N^2}
	\left(\frac{1}{p^2_{K-1}} + \delta_{ij} \frac{1}{p_i^2} \right) \text{with } i, j
	\in [0, ..., K-2],
\end{align}
which differs from the expression for fluorescence in the $N$-dependent factor as well as 
the square on $p_i$. The result in terms of position, $F_{\bar{r}_m}$, is given by
\begin{equation}
	F_{\bar{r}_m} = {\mathcal{J}^*}^{\top} F_{\bar{p}} \mathcal{J}^*,
\end{equation}
with $\mathcal{J}^* \in \mathbb{R}^{(K-1) \times d}$ the Jacobian matrix of the transformation 
from $\bar{r}$-space to the reduced $\bar{p}$-space, which is a $(K-1)$-dimensional space 
since $p_{K-1} = 1 - \sum^{K-2}_{j=0} p_j$. There are, therefore, $K-1$ \textit{independent} 
parameters. We finally get
\begin{equation}
	\label{fishinfeq}
	F_{\bar{r}_m} = \frac{\hat{N}^2_{int} - \sigma_N^2}{\sigma_N^2} \sum^{K-1}_{i=0}
	\frac{1}{p_i^2} 
	\begin{bmatrix}
		\left(\frac{\partial p_i}{\partial r_{m1}}\right)^2 & \dots &
		\frac{\partial p_i}{\partial r_{m1}} \frac{\partial p_i}{\partial
		r_{md}}\\ 
		\vdots & \ddots & \vdots\\
		\frac{\partial p_i}{\partial r_{md}} \frac{\partial p_i}{\partial
		r_{m1}}& \dots &\left(\frac{\partial p_i}{\partial r_{md}}\right)^2
		 
	\end{bmatrix}.
\end{equation}
The Cramér-Rao lower bound on the covariance matrix $\Sigma (\bar{r}_m)$ is given by
$\Sigma_{CRB} = F_{\bar{r}_m}^{-1}$. In our results, we use the arithmetic mean of the
eigenvalues of this matrix (recall that $d$ is the number of dimensions):
\begin{equation}
\label{sigmacrb}
\tilde{\sigma}_{CRB} = \sqrt{\frac{1}{d} 
\tr(\Sigma_{CRB})}.
\end{equation}

\section{Results}
\label{sec:results}

\subsection{Static localization}
\subsubsection{Fluorescence}

\begin{figure*}[!htp] 
    \centering
	\begin{subfigure}[t]{\textwidth}
        \center{\includegraphics{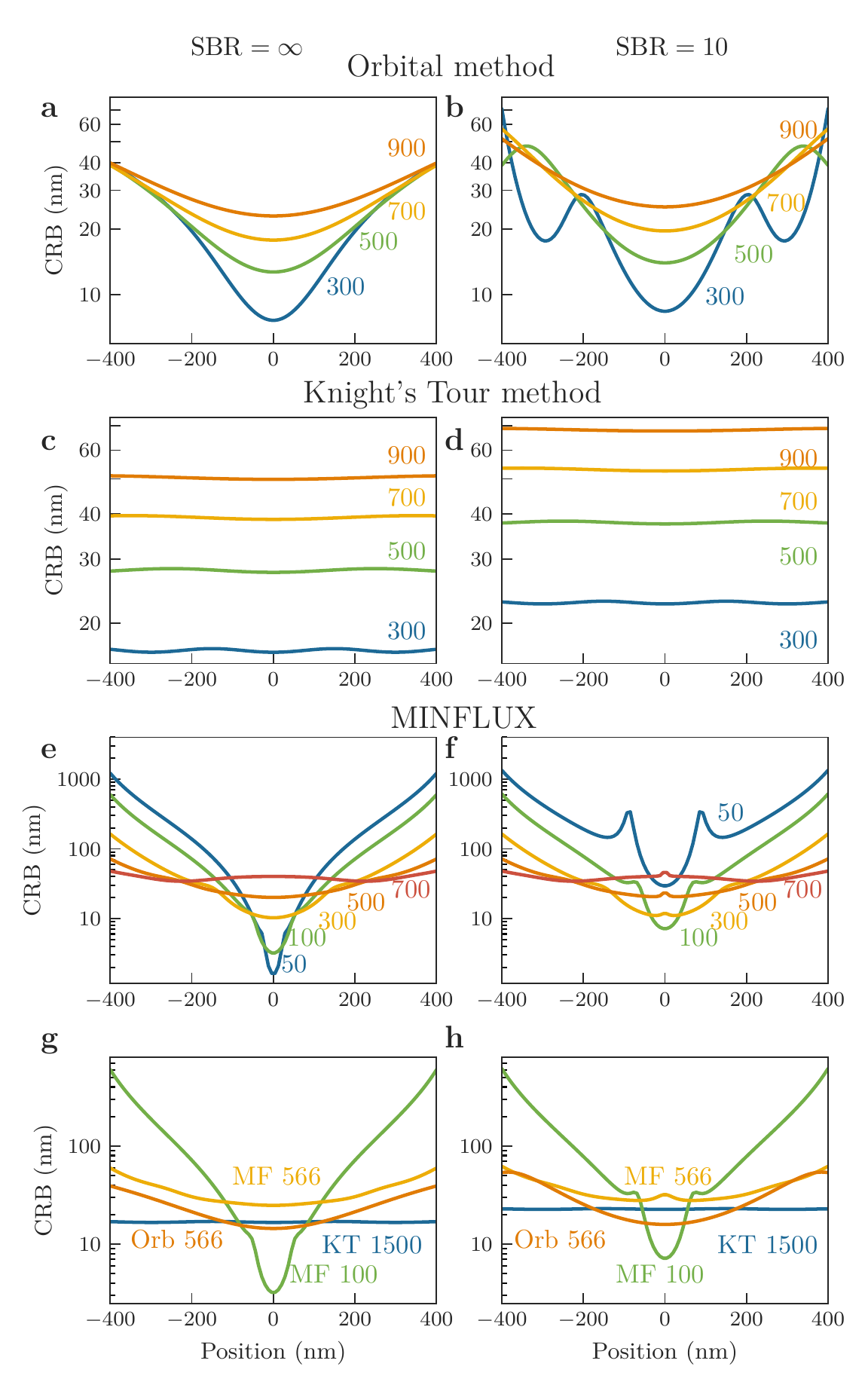}}
         \phantomcaption \label{crbpanel:a}   
    \end{subfigure}
	\begin{subfigure}[t]{0\textwidth}
         \includegraphics[width=\textwidth]{crb_panel.pdf}
         \phantomcaption \label{crbpanel:b}   
    \end{subfigure}
	\begin{subfigure}[t]{0\textwidth}
         \includegraphics[width=\textwidth]{crb_panel.pdf}
         \phantomcaption \label{crbpanel:c}   
    \end{subfigure}
	\begin{subfigure}[t]{0\textwidth}
         \includegraphics[width=\textwidth]{crb_panel.pdf}
         \phantomcaption \label{crbpanel:d}   
    \end{subfigure}
	\begin{subfigure}[t]{0\textwidth}
         \includegraphics[width=\textwidth]{crb_panel.pdf}
         \phantomcaption \label{crbpanel:e}   
    \end{subfigure}
	\begin{subfigure}[t]{0\textwidth}
         \includegraphics[width=\textwidth]{crb_panel.pdf}
         \phantomcaption \label{crbpanel:f}   
    \end{subfigure}
	\begin{subfigure}[t]{0\textwidth}
         \includegraphics[width=\textwidth]{crb_panel.pdf}
         \phantomcaption \label{crbpanel:g}   
    \end{subfigure}
	\begin{subfigure}[t]{0\textwidth}
         \includegraphics[width=\textwidth]{crb_panel.pdf}
         \phantomcaption \label{crbpanel:h}   
    \end{subfigure}
	\caption{\label{crbpanel} CRB along the x-axis for different $L$-values (indicated in nm 
	next to each curve) for (a, b) 
	orbital method, (c, d) Knight's Tour method, (e, f) MINFLUX, (g, h) different methods 
	(MF - MINFLUX, KT - Knight's Tour, Orb - orbital method). 
	$N=100$ photons for all curves. Left column: SBR $=\infty$; right column: SBR~$=10$. 
	For MINFLUX, SBR~$=10$ corresponds with $L=\SI{100}{\nano\meter}$, while SBR was scaled for 
	other $L$-values (see text for details).}
\end{figure*}

We first compare the three different scan patterns based on fluorescence using $N=100$
photons at two different signal-to-background ratio (SBR) values and different values of
the scanning length $L$.  For the orbital and Knight's Tour methods, the beam waist was
scaled relative to $L$. For the orbital method, the relation $L = 2w/\sqrt{2}$ was used,
while for the Knight's Tour method, the grid spacing was chosen such that $L = 3.75w$, 
consistent with Ref.~\onlinecite{wangOptimalStrategyTrapping2010}.
Therefore, for these two methods, $L$ is bounded by the diffraction limit, whereas for
MINFLUX it is not.  The scaling of the beam size with scanning length for the orbital and
Knight's Tour methods also means that the SBR does not depend on the scanning length for these methods 
(if the beam size is kept constant, an increase in the scanning length would lead to a decrease in the SBR). For 
MINFLUX, however, there is a dependence of SBR on scanning length; specifically, the SBR scales with the
average illumination intensity and thus
\textit{increases} with increased $L$. In this case, the assumed SBR of 10 was for 
$L=\SI{100}{\nano\meter}$, and the adjusted SBR for other scanning lengths ranged from 2.5 for 
$L=\SI{50}{\nano\meter}$ to 300 for $L=\SI{700}{\nano\meter}$. In practice, the SBR would be bounded by 
saturation, but since MINFLUX does not show good performance at large $L$, these results are not of practical relevance. Furthermore, at these large $L$ values, the results for different values of the SBR~$\geq 10$ are very similar to each other and the results in \cref{crb_iscat:f} are therefore representative of the performance.  The results 
are shown in \cref{crbpanel}.

All three methods show improved precision with decreased $L$, with the lower SBR showing a 
general decrease in precision. The differences between the three methods are also clear. The 
orbital method (\cref{crbpanel:a,crbpanel:b}) shows a lower CRB
in the center of the scan range, along with a moderate decrease in CRB with decreased
$L$. The Knight's Tour method (\cref{crbpanel:c,crbpanel:d}) has a very even CRB across the 
scan range due to even
illumination. It too has a moderate improvement in precision with decreased $L$. MINFLUX
(\cref{crbpanel:e,crbpanel:f}) shows a dramatically lower CRB in the center of the scan range 
compared to the edges, along with a large
decrease in the central CRB with decreased $L$. Here we also see the greatest impact of
the SBR: in the case of SBR$ =\infty$ (\cref{crbpanel:e}), the central CRB decreases 
without bound as $L$ tends to 0, whereas with a finite SBR (\cref{crbpanel:f}) there is a limit 
to how low the central CRB can be made by decreasing $L$. The three methods are compared 
directly in \cref{crbpanel:g,crbpanel:h} where the beam waist for the orbital and Knight's 
Tour methods is set to $w=400$~nm, and two different
MINFLUX $L$-values are shown. It is clear that MINFLUX indeed provides more information
per photon, but the trade-off is a smaller scanning range. With the same value of $L$, the
performance of MINFLUX is somewhat worse than that of the orbital method. This fact is
important, as a large scanning range is required for tracking fast-moving particles.

\subsubsection{iSCAT}
\label{iscatacc}

\begin{figure*}[!htp] 
	\begin{subfigure}[t]{\textwidth}
        \center{\includegraphics{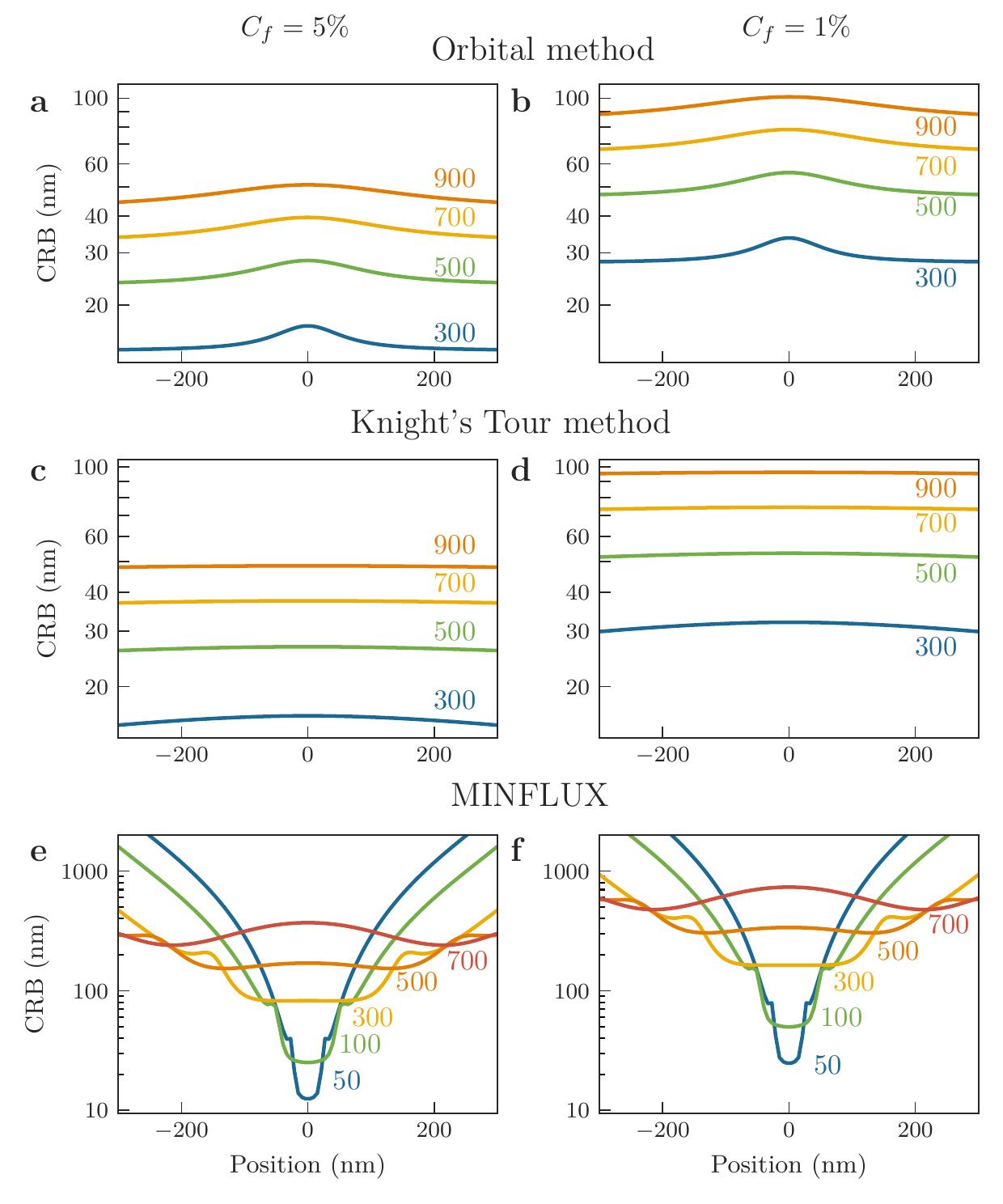}}
         \phantomcaption \label{crb_iscat:a}   
    \end{subfigure}
	\begin{subfigure}[t]{0\textwidth}
         \includegraphics[width=\textwidth]{crb_iscat_panel.pdf}
         \phantomcaption \label{crb_iscat:b}   
    \end{subfigure}
	\begin{subfigure}[t]{0\textwidth}
         \includegraphics[width=\textwidth]{crb_iscat_panel.pdf}
         \phantomcaption \label{crb_iscat:c}   
    \end{subfigure}
	\begin{subfigure}[t]{0\textwidth}
         \includegraphics[width=\textwidth]{crb_iscat_panel.pdf}
         \phantomcaption \label{crb_iscat:d}   
    \end{subfigure}
	\begin{subfigure}[t]{0\textwidth}
         \includegraphics[width=\textwidth]{crb_iscat_panel.pdf}
         \phantomcaption \label{crb_iscat:e}   
    \end{subfigure}
	\begin{subfigure}[t]{0\textwidth}
         \includegraphics[width=\textwidth]{crb_iscat_panel.pdf}
         \phantomcaption \label{crb_iscat:f}   
    \end{subfigure}
	\caption{\label{crb_iscat} CRB along the x-axis for different $L$-values (indicated in nm 
	next to each curve) for (a, b) 
	orbital method, (c, d) Knight's Tour method and (e, f) MINFLUX with iSCAT localization. 
	$N_{int}=100$ photons, left column: $C_f=5\%$, right column: $C_f=1\%$.}
\end{figure*}

We next compare the different scan patterns using iSCAT-based localization, considering 
$N_{int} = 100$ and two contrast values: $C_f = 5\%$ and $C_f = 1\%$ (\cref{crb_iscat}). 
The shape
of the CRB curve for the orbital method (\cref{crb_iscat:a,crb_iscat:b}) is quite different 
than with fluorescence. As we
have previously shown~\cite{vanheerdenTheoreticalComparisonRealtime2021b}, this is due to
the uncertainty in the parameter $p_i$ (Eq.~\ref{parameq}) having a maximum in the center, both
for iSCAT and fluorescence. For fluorescence, this uncertainty is counteracted by the 
dependence of the position uncertainty
on the derivative of $p_i$, which has a minimum in the center. 
For iSCAT, the dependence on the parameter uncertainty itself is quadratic and therefore
weighs more strongly. The CRB curve for the Knight's Tour method (\cref{crb_iscat:c,crb_iscat:d}) is 
again very flat. MINFLUX (\cref{crb_iscat:e,crb_iscat:f}) shows a similar trend as with 
fluorescence, with the central precision again depending strongly on the scanning length, 
resulting in a good precision only in the center. 

\begin{figure}[!htp]
    \centering
    \includegraphics{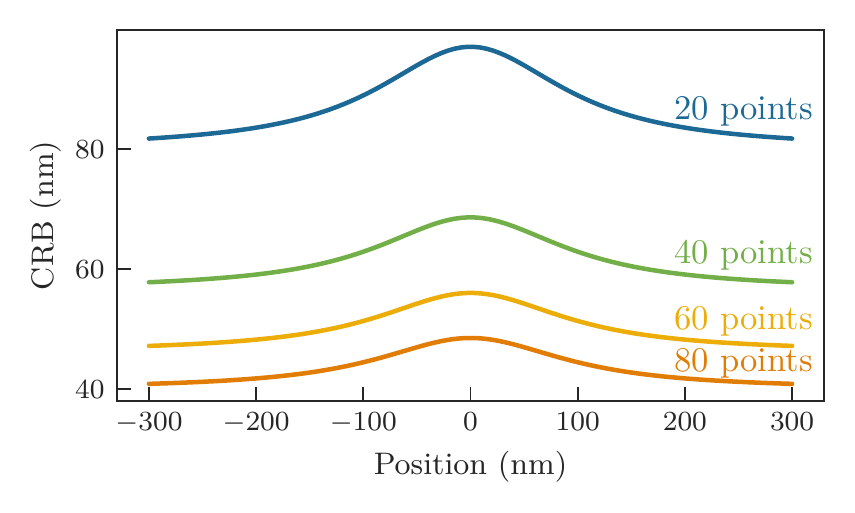}
    \caption{Influence of the number of scan points on the iSCAT CRB for the orbital method.
    For all curves, $L=\SI{500}{\nano\meter}$, $N_{int} = 100$ photons, $C_f = 1\%$.}
    \label{numpoints}
\end{figure}

An interesting feature of the iSCAT CRB expression is its dependence on the number of scan 
points (\cref{numpoints}). The improvement in precision when using more scan points is due to a 
more accurate measurement of the \textit{background} when it is measured at every point.

In order to compare iSCAT and fluorescence, we consider the relative number of
fluorescence and scattering counts at the same illumination intensity, as well as the
average iSCAT contrast, for a specific sample. The average contrast is proportional to
$2\sqrt{\nicefrac{\sigma_{scat}}{r}}$, where $\sigma_{scat}$ is the scattering
cross-section and $r$ is the reflectivity of the interface, while the relative fraction
of iSCAT counts to fluorescence counts, $\nicefrac{N_{scat}}{N}$, is proportional to
$\nicefrac{2\sqrt{\sigma_{scat}r}}{\sigma_{abs}\Phi_f}$, where $\sigma_{abs}$ and
$\Phi_f$ are the absorption cross-section and fluorescence quantum yield, respectively.
We consider four example samples that may be tracked using both methods: an HIV virion 
labeled with the quantum dot QD525, the eGFP mutant of green fluorescent protein 
(a mutant featuring enhanced fluorescence), the main light-harvesting complex of plants 
(LHCII), and phycobilisome (PB), the main light-harvesting from the cyanobacterial strain \textit{Synechocystis} PCC6803. These examples include intrinsically fluorescent (eGFP, LHCII and PB) and 
artificially labeled (QD525-labeled HIV) nanoscaled biological systems displaying a 
wide range of sizes. GFP is a relatively small protein, while PB is a very large 
(6.2 MDa~\cite{sauerStructuresCyanobacterialPhycobilisome2021}) multidomain 
pigment-protein complex; LHCII is a trimeric pigment-protein complex with a molecular 
weight of 128 kDa~\cite{panStructuralBasisLhcbM5mediated2021}, while HIV is an order of 
magnitude larger in all dimensions. Furthermore, GFP and PB are both water-soluble, while LHCII is a membrane
protein and typically embedded in detergent micelles for \textit{in vitro} studies. 

The scattering cross-section of a particle can be estimated using the
formula $\sigma_{scat} = \frac{8}{3}\pi^3 \alpha^2 (\lambda / n_m)^{-4}$, where $\alpha =
3V (n_s^2 - n_m^2)/(n_s^2 + 2n_m^2)$ is the particle's polarizability, $V$ is its volume,
$n_s$ its refractive index, $n_m$ the refractive index of the surrounding medium, and
$\lambda$ the wavelength of the incident light. The calculated values for
$\sigma_{scat}$, $C_f$, and $\nicefrac{N_{scat}}{N}$ are displayed in
Table~S1, along with other relevant parameters derived from literature. 
$C_f$ is calculated from \cref{conteq}. 
We assume HIV to be spherical with a radius of 73 nm. PB is hemidiscoidal with a
radius of 30 nm and a thickness of 10 nm. LHCII is cylindrical in shape with a radius of
3.65 nm and a  height of 4.5 nm. GFP is also cylindrical with a radius of 1.2 nm and a
height of 4.2 nm. We also consider the case of LHCII in a detergent micelle, as this is
typically how it is measured in solution. In this case, the length of an
n-dodecyl-\textbeta-D-maltoside molecule (2.3~nm) was added to the effective radius. In
each case, the effective radius is given in Table~S1, equal to the
radius of a sphere with the same volume as the particle. The peak absorption wavelength
(except for QD525 for which the absorbance continues to increase at shorter wavelengths)
of each sample is also shown in Table~S1. The reflectivity of a
glass-water interface at normal incidence (0.4\%) for a typical microscope coverslip was used for the calculations.

\begin{figure*}[!htp] 
    \centering
	\begin{subfigure}[t]{\textwidth}
        \center{\includegraphics{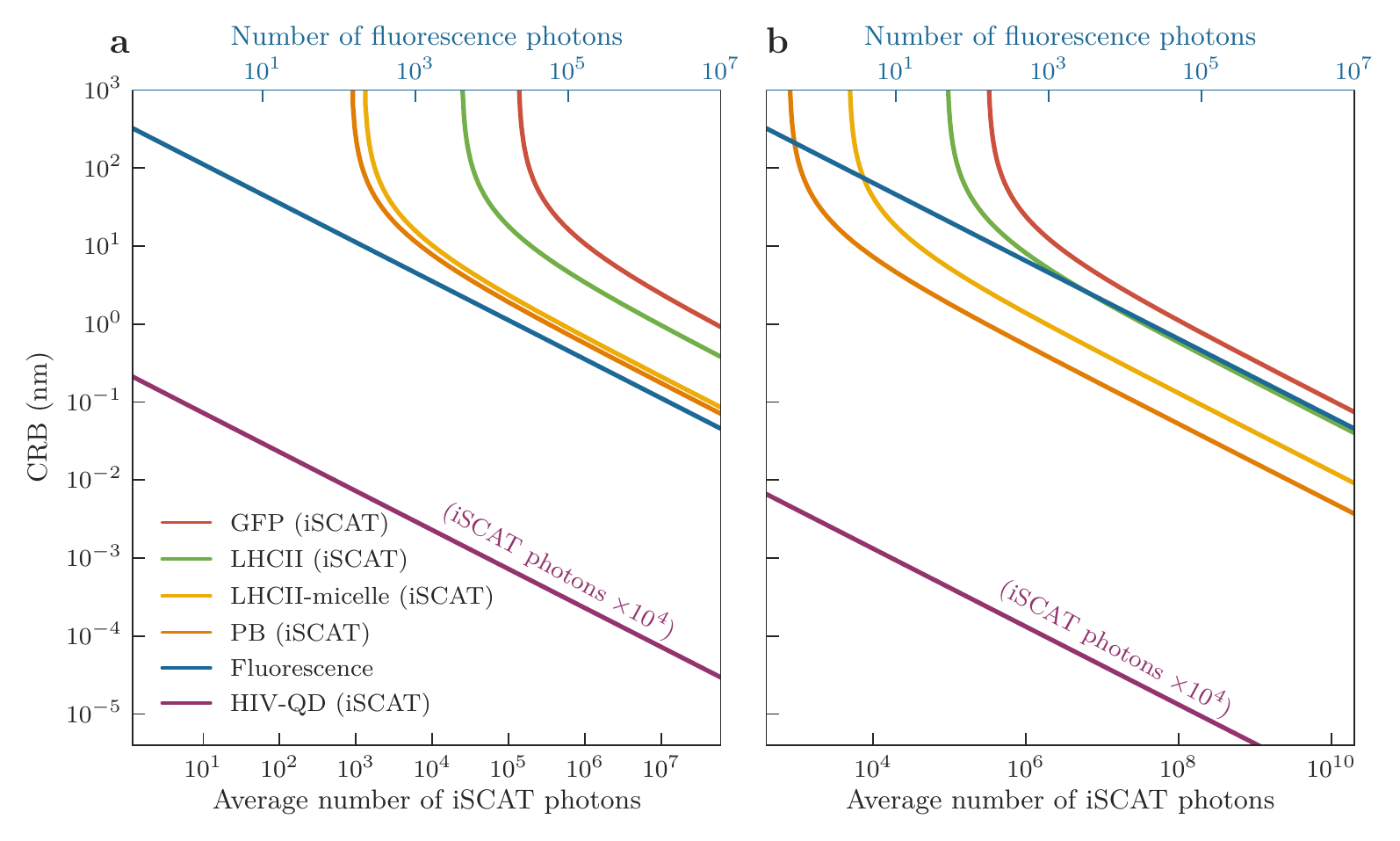}}
         \phantomcaption \label{iscat_numphotons:a}   
    \end{subfigure}
	\begin{subfigure}[t]{0\textwidth}
         \includegraphics[width=\textwidth]{comp_numphotons.pdf}
         \phantomcaption \label{iscat_numphotons:b}   
    \end{subfigure}
	\caption{\label{iscat_numphotons} CRB as a function of the number of photons for
	both iSCAT and fluorescence for different samples, using the same illumination
	intensity for \textbf{a} and a 1000 times increased intensity for iSCAT for \textbf{b}. The upper x-axes 
	show fluorescence counts. The iSCAT counts are higher than 
	the fluorescence counts by the
	ratios given in \cref{example_table,example_table_2} for \textbf{a} and \textbf{b}, respectively. 
	The average 
	iSCAT counts for all samples except HIV-QD are shown on the lower x-axes. For HIV-QD, the iSCAT
	counts are ${\sim} 10^4$ times higher than the average for the other samples, as 
	indicated next to the curve.}
\end{figure*}

The CRB as a function of the detected number of photons is shown in
\cref{iscat_numphotons} for iSCAT and fluorescence. The iSCAT performance follows the
size of the particle. For a large particle such as a virus, iSCAT is orders of magnitude
more precise than fluorescence, whereas for small particles such as GFP or LHCII,
fluorescence-based localization is better. For LHCII in a micelle and PB, the precision
of the two methods is similar. The asymptotic shape of the iSCAT curve is due to the factor
$\nicefrac{\hat{N}^2 - \sigma_N^2}{\sigma_N^2}$ in the Fisher information expression 
(Eq.~\ref{fishinfeq}). The corresponding factor for fluorescence is simply $\hat{N}$
\cite{balzarottiNanometerResolutionImaging2017}. For large values of $\hat{N}$,
$\nicefrac{\hat{N}^2 - \sigma_N^2}{\sigma_N^2}$ is approximately equal to
$\nicefrac{\hat{N}^2}{\sigma_N^2}$. If we assume $\hat{N}$ to be large and Poisson-distributed, then $\sigma_N^2 \approx \hat{N}$ and therefore 
$\nicefrac{\hat{N}^2}{\sigma_N^2}
\approx \hat{N}$. It follows that for both iSCAT and fluorescence, at high count rates, 
$\tilde{\sigma}_{CRB} \propto \hat{N}^{-\frac{1}{2}}$  (Eq.~\ref{sigmacrb}). 

The result in \cref{iscat_numphotons:a} assumes the same illumination intensity for
iSCAT and fluorescence. However, iSCAT can be performed at a wavelength where the sample
does not strongly absorb, and thus it is possible to use a much higher illumination
intensity for iSCAT at a wavelength where phototoxicity is strongly limited, leading to improved precision. To investigate this possibility, we
chose a separate wavelength for iSCAT for each sample, and assumed a 1000 times increased
illumination intensity. This is based on single-molecule fluorescence experiments
typically using an excitation power of around 1~\textmu W, and iSCAT experiments using around
1~mW.  The calculation values in this case are shown in Table~S2 and the
results are shown in \cref{iscat_numphotons:b}. In this case,
both PB and LHCII in a micelle have a markedly improved precision with iSCAT, and for
pure LHCII and GFP the precision is roughly the same for the two methods.

\begin{figure*}[htp] 
	\center{\includegraphics{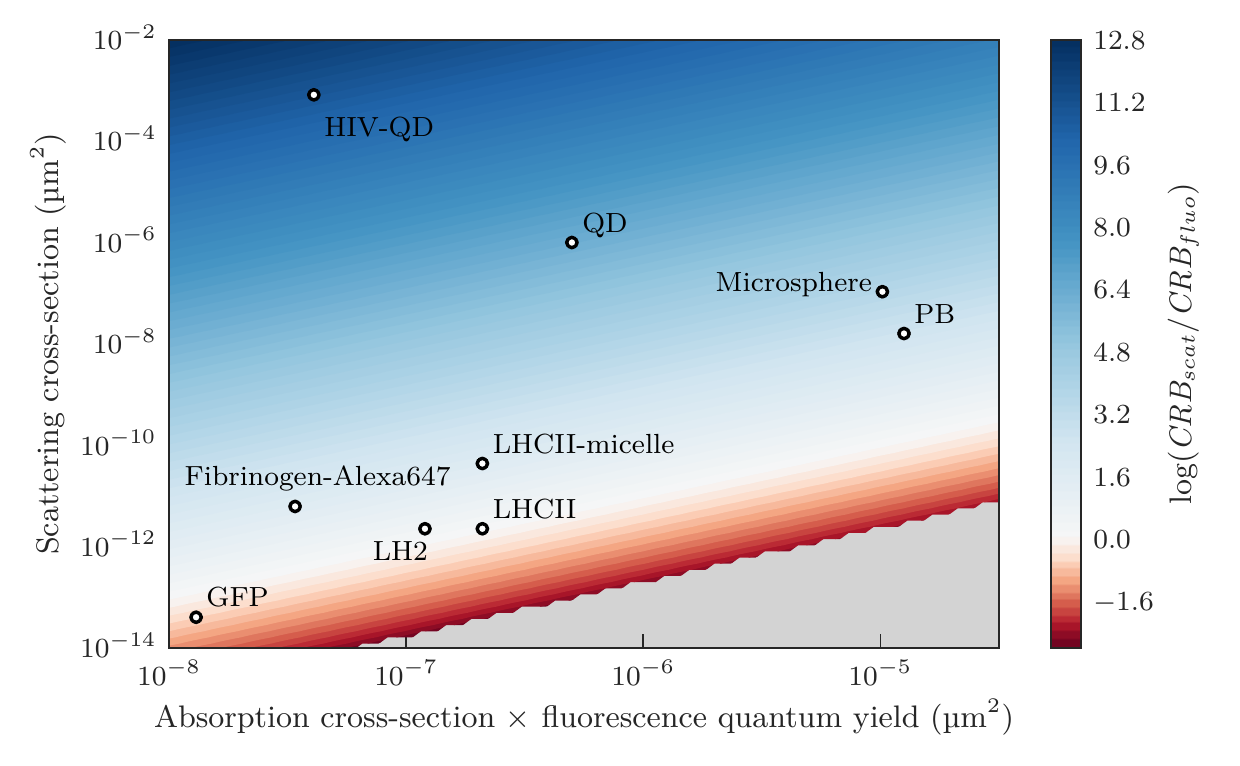}}
	\caption{\label{2dcrb} Logarithm of the ratio of iSCAT CRB to fluorescence CRB as a function 
	of the scattering cross-section and the product of absorption cross-section with fluorescence 
	quantum yield. $N = \SI{1e4}{}$ 
	fluorescence photons and the iSCAT illumination is assumed to be 1000
    times higher than the fluorescence illumination (each sample has a different assumed illumination that would produce $\SI{1e4}{}$ fluorescence photons). Positive (blue) values indicate better 
    precision with iSCAT while negative
    (red) values indicate better precision with fluorescence. The positions of the example 
    samples are indicated. The lower right region is excluded due to undefined values 
    (corresponding to low numbers of photons in \cref{iscat_numphotons}).}
\end{figure*}

The relationship between scattering and absorption cross-sections
and the iSCAT and fluorescence CRBs is further illustrated in \cref{2dcrb}. Here the logarithm of 
the ratio of iSCAT CRB to fluorescence CRB is depicted as a function of the scattering 
cross-section and the product of the absorption cross-section with the fluorescence quantum yield. 
The examples considered above are indicated, along with a number of other representative samples: 
a 12-nm diameter CdSe/ZnS quantum dot (QD), a 40-nm fluorescent microsphere, fibrinogen labeled
with Alexa Fluor 647, and the bacterial light harvesting complex LH2. These samples all show 
improved precision with iSCAT, to varying degrees. It is a simple task to estimate the position of 
any other sample, and thus evaluate the possible suitability of iSCAT for that sample, using this 
figure.

\subsection{Dynamic simulation}
\subsubsection{Control System Design} To compare the methods' dynamic performance, the
complete control system was simulated numerically. The output is assumed to be a piezoelectric
sample stage, which is assumed to have a linear
response, a general feature of piezoelectrics. For simplicity we design a second-order 
system with a step response similar to that of a real stage (with its manufacturer-provided 
controller) (\cref{step}). Details can be found in the SI; we show the final result here. 
With the system state
\begin{align}
	\vec{x} = \begin{bmatrix}
		x_p \\
		x_s \\
		\dot{x_s}
	\end{bmatrix},
\end{align}
where $x_p$ and $x_s$ are the particle and stage positions, respectively, the dynamical system is 
given by
\begin{align}
	\vec{\dot{x}} &= \begin{bmatrix}
		0 & 0 & 0\\
		0 & -2 & 4\\
		0 & -4 & 0
	\end{bmatrix} \vec{x} +
	\begin{bmatrix}
		0 \\
		2 \\
		4
	\end{bmatrix} u +
	\begin{bmatrix}
		\sqrt{2D}\\
		0\\
		0
	\end{bmatrix} \xi(t)\\
	y &= \begin{bmatrix}
		1 & -1 & 0 \\
	\end{bmatrix} \vec{x} + w_n,
	\label{sysode}
\end{align}
where $u$ is the control input, $D$ is the two-dimensional diffusion coefficient, $\xi$ is a 
white-noise process
with mean 0 and variance 1, $w_n$ is an error term that mostly originates from
Poisson noise propagating through the measurement process, and $y$ is the measured position.
We also implemented a Kalman filter as this is
commonly used in real RT-FD-SPT systems.

For the orbital method, we used the position estimate from Ref.~\onlinecite{berglundFeedbackControllerDesign2004}, for MINFLUX, we used the online position estimator from Ref.~\onlinecite{balzarottiNanometerResolutionImaging2017}, and for the Knight's Tour method, we used 
the estimator
\begin{align} 
\hat{\bar{r}} = \sum^{40}_{i=1}\hat{p}_i \cdot \bar{r}_{b_i},
\end{align}
where $\hat{\bar{r}}$ is the estimated position vector, $\bar{r}_{b_i}$ are the position 
vectors of the beam positions $b_i$, and $\hat{p}_i$ are the estimated values of the
parameters defined in Eq.~\ref{parameq}, based on the detected photons at each scan point.

We performed dynamic simulations of each method by
numerically integrating Eq.~\ref{sysode} in two dimensions, computing estimates for 
$\vec{y}_k$ using the estimators mentioned above, and subsequently estimating $\vec{x}_k$ 
using the Kalman filter equations ($\vec{y}_k$ and $\vec{x}_k$ are the discretization of 
$\vec{y} = y$ and 
$\vec{x}$ respectively --- see the SI for details). The estimated position $x_p$ was directly 
applied as
feedback into the control input $u$.

The parameters used were as follows: the beam waist for the orbital and Knight's Tour 
methods was $w=\SI{400}{\nano\meter}$, the MINFLUX beam size parameter was 
$\mathrm{fwhm}=\SI{500}{\nano\meter}$, the MINFLUX scanning length was 
$L=\SI{50}{\nano\meter}$, the detected count rate for a particle in the center of the beam 
was $\Gamma_0=12.5$ kcounts/s, and the fluorescence signal-to-background ratio was 10. For
iSCAT-based localization, the same assumptions were made for the average contrast and ratio 
of iSCAT to fluorescence counts as in the case of static
precision. The beam scan was performed at \SI{8.33}{\kilo\hertz}, 
and feedback was applied at the same frequency. The Kalman noise covariance $R$ was
adjusted empirically for each scan pattern and iSCAT sample, to achieve the fastest possible 
tracking.

\begin{figure*}[htp]
	\center{\includegraphics{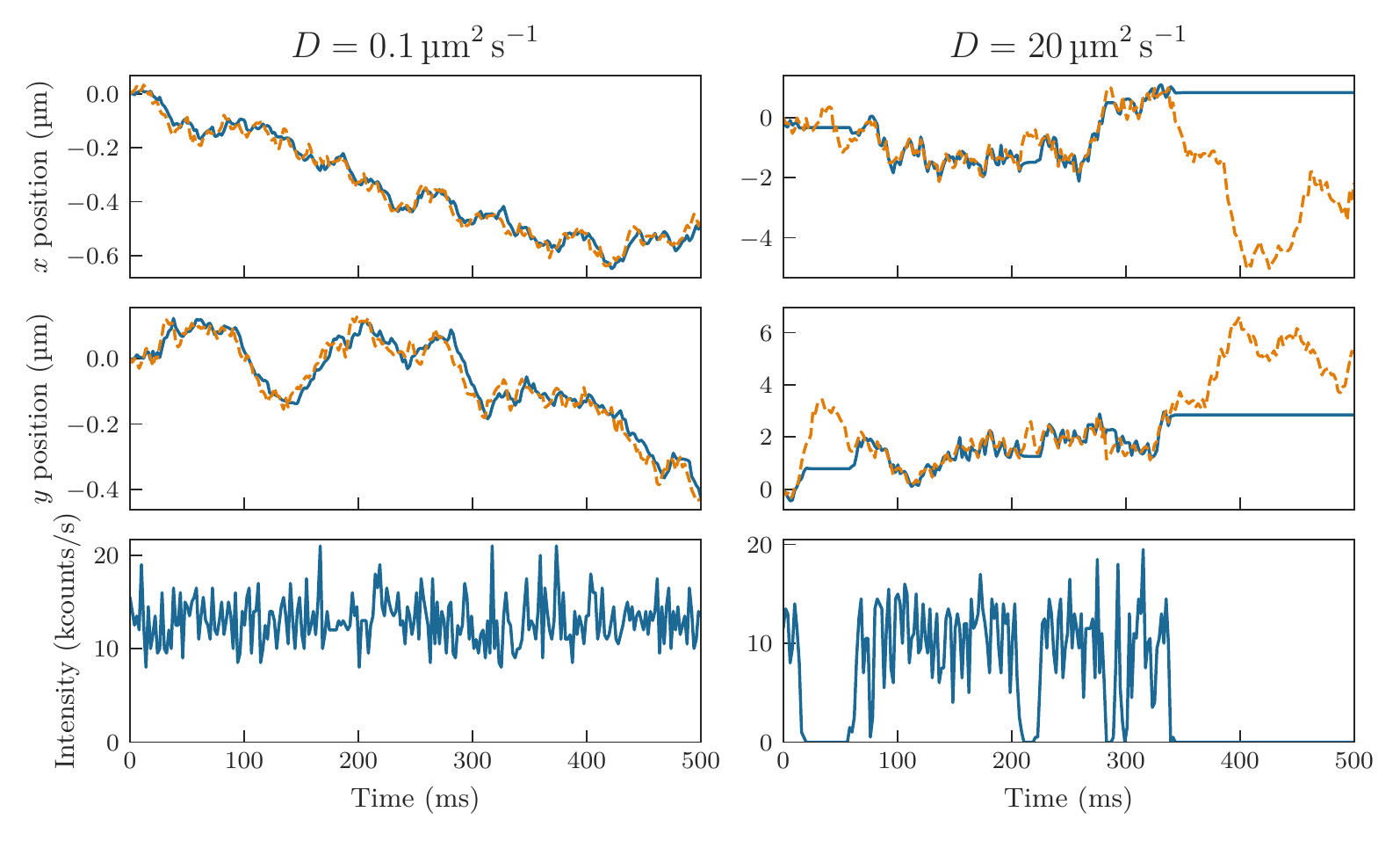}}
	\caption{\label{trajgood} Example of a well-tracked trajectory (left) and an "almost tracked" trajectory (right). The dashed orange curves show the particle position, while 
	the blue solid curves shows the 
	stage position. Note the difference in scale for the position axes, and the large jumps 
	in the intensity for the trajectory on the right.}
\end{figure*}

\begin{figure*}[htp]
	\begin{subfigure}[t]{\textwidth}
        \center{\includegraphics{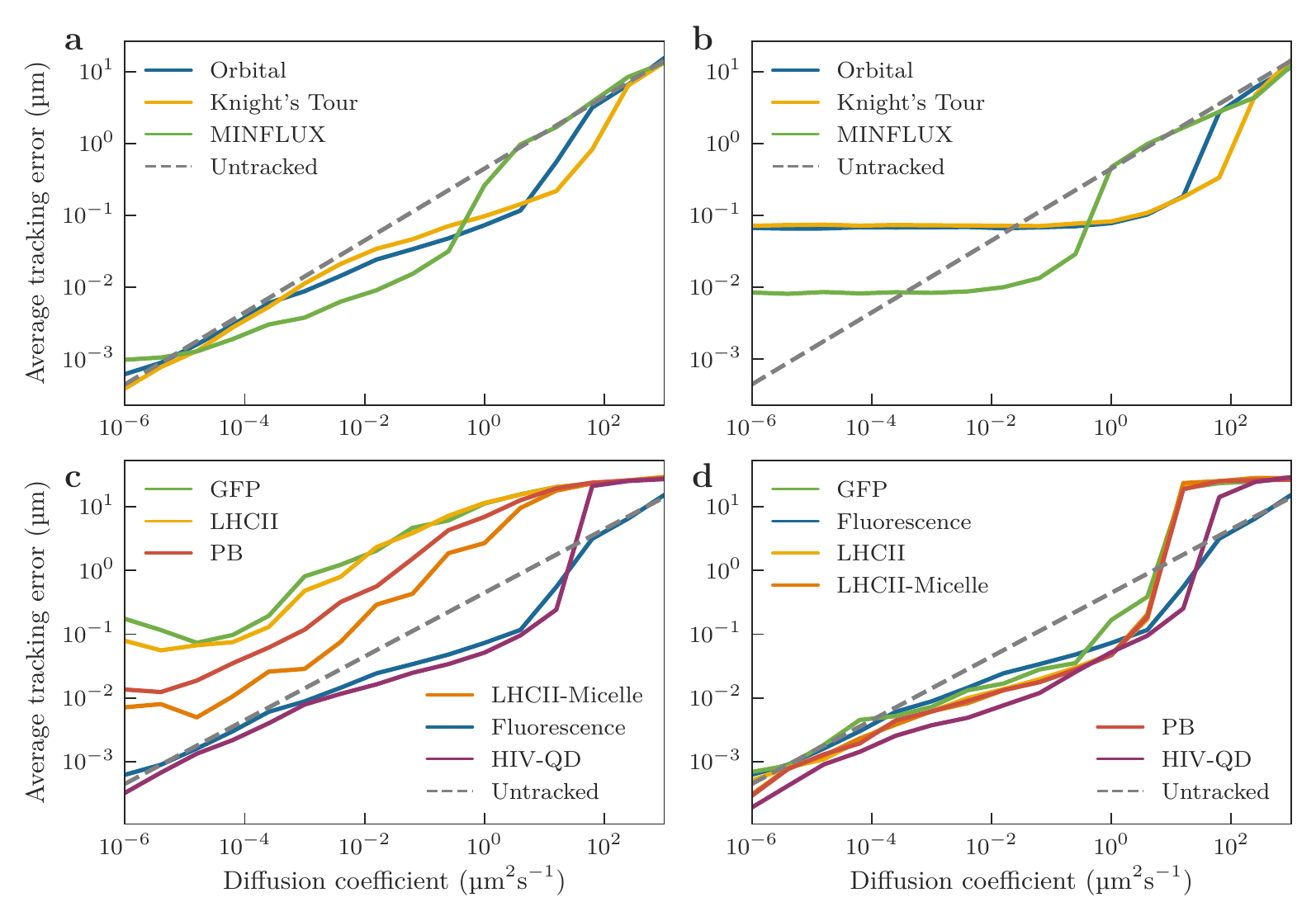}}
         \phantomcaption \label{methoderr:a}   
    \end{subfigure}
	\begin{subfigure}[t]{0\textwidth}
         \includegraphics[width=\textwidth]{err_diff_combined.pdf}
         \phantomcaption \label{methoderr:b}   
    \end{subfigure}
	\begin{subfigure}[t]{0\textwidth}
         \includegraphics[width=\textwidth]{err_diff_combined.pdf}
         \phantomcaption \label{methoderr:c}   
    \end{subfigure}
	\begin{subfigure}[t]{0\textwidth}
         \includegraphics[width=\textwidth]{err_diff_combined.pdf}
         \phantomcaption \label{methoderr:d}   
    \end{subfigure}
	\caption{\label{methoderr} Average tracking error as a function of the diffusion 
	coefficient for (a,b) the three different scanning patterns using fluorescence-based 
	localization (a) with and (b) without a Kalman filter, and (c,d) different samples using
	iSCAT and fluorescence with (c) the same illumination intensity and (d) a 1000 increased illumination intensity for iSCAT. A constant fluorescence intensity was
	assumed for all samples, justifying why (c) and (d) show only one curve for 
	fluorescence. For (a), (b), and (d), an average over 5 runs was used for each value of the
	diffusion coefficient, while for (c) the average was over 100 runs due to the large
	variance in the average error of the unsuccessfully tracked samples. The dashed lines labeled "untracked" represent the expected error for a freely diffusing particle.}
\end{figure*}

A typical well-tracked trajectory is 
shown in the left column of \cref{trajgood}, where the diffusion coefficient is 
\SI{0.1}{\micro\metre\squared\per\second}.
The stage closely follows the diffusion and the intensity shows only Poisson noise. An
"almost tracked" trajectory is shown in the right column of \cref{trajgood} where the diffusion
coefficient is \SI{20}{\micro\metre\squared\per\second}. Here, the system is unable to
respond fast enough. There are large jumps in the measured intensity as the
particle moves in and out of the illumination beam before eventually being lost completely. 
Performing many such simulation runs
for different values of $D$, we can calculate the average tracking error, and the results 
are shown in \cref{methoderr}. 

We first compare the different scanning patterns using fluorescence as shown in 
\cref{methoderr:a}. For each method, three different regions can be identified. At high values 
of $D$, there is an "untracked" region, where
the tracking error follows free particle statistics, i.e., $e_{tau} = \sqrt{2D\tau}$, indicated
by the dashed line. At intermediate values of $D$, there is a "well-tracked" region, where 
the error is lower than the free particle value. At the lowest values of $D$, the system
starts tracking noise and the error is worse than in the case when no tracking takes place. 
Another way of looking at the latter regime is to say that the feedback frequency is too 
fast~\cite{berglundPerformanceBoundsSingleparticle2006}. 

The scanning methods give rise to at least two important differences. Firstly, in the 
well-tracked region, the precision of the tracking error lines up with the static
result (\cref{crbpanel}), with MINFLUX having the best precision, and the Knight's
Tour method having the worst precision. Specifically, the precision in the center of the
tracking region determines the tracking error, since good tracking keeps the particle
close to the center. Secondly, each method has a different "cutoff" value where the
diffusion becomes too fast to track. This value is determined by the size of the tracking 
area and also confirms the static result. MINFLUX has the lowest cutoff due to its small 
tracking range ($L=\SI{50}{\nano\meter}$), followed by the orbital 
($L=\SI{566}{\nano\meter}$) and Knight's Tour ($L=\SI{1500}{\nano\meter}$) methods. For very 
slow diffusion, it is clear that the precision does not follow the same trend as in the 
faster tracked region. This is due to different values being used for the Kalman covariance 
$R$ for the different scanning patterns. These values were optimized to achieve the
highest possible "cutoff" value for each method, and the Knight's Tour has the largest 
value. As a result, the effect of noise is dampened, artificially providing better 
performance in this "noise-tracking" regime. In a real experiment, one would simply decrease 
the feedback frequency for tracking slower diffusion. 

A comparison between iSCAT and fluorescence for the different example samples is shown in 
\cref{methoderr:c,methoderr:d}. For simplicity we assume a constant fluorescence intensity for 
all the samples, since we are only interested in the relative
performance of iSCAT vs fluorescence-based localization. Hence, there is only one curve
for fluorescence. The general trend is the same as in \cref{iscat_numphotons}, and the 
main result of static precision is confirmed: for all samples except GFP, iSCAT performs
better than fluorescence when an increased illumination intensity is used. Interestingly, the 
improvement over fluorescence is even better
here, with even LHCII showing markedly better tracking using iSCAT.

One noticeable difference between the dynamic tracking precision and static CRB is the smaller difference in precision between different methods and samples in the case of dynamic tracking. This is mostly due to the Kalman filter successfully suppressing the
influence of noise. In \cref{methoderr:b}, the tracking error is shown for the different
scanning patterns using a simulation without a Kalman filter. To amplify the results, a higher 
fluorescence intensity (60 kcounts/s) was used. Here, the tracking error matches the CRB more 
closely. Comparing the result with and without the Kalman filter
highlights the power of the Kalman filter to compensate for noise. However, there are 
still fundamental limits to precision that the Kalman filter is unable to fully overcome.

\section{Discussion and Conclusion}
\label{sec:dissconc}

The results for the different scanning patterns highlight the fundamental trade-off between 
tracking speed and tracking precision. The orbital method was shown to provide optimal
position information when using a Gaussian beam~\cite{gallatinOptimalLaserScan2012}. The 
success of MINFLUX demonstrates, however, that much better precision can be achieved with
a non-Gaussian beam profile --- specifically, one with an illumination minimum. The Knight's Tour 
method, on the other hand, while having worse precision than
the orbital method, is much faster due to a much larger scan area. The 
Knight's Tour method (or at least a three-dimensional version of it, 3D-DyPLoT) has also shown better experimental results than the orbital method in terms of RT-FD-SPT. While both methods
have been shown experimentally to track particles up to $D=\SI{20}{\micro\meter\squared\per\second}$ (Ref. ~\onlinecite{mchaleQuantumDotPhoton2007, houRobustRealtime3D2017}), 3D-DyPLoT has achieved this at a 6 times lower count rate (\SI{20}{\kilo\hertz} vs \SI{120}{\kilo\hertz}~\cite{mchaleQuantumDotPhoton2007, houRobustRealtime3D2017}).

It has recently been shown theoretically and experimentally that selectively sampling only 
some locations in the Knight's Tour
pattern results in improved precision~\cite{zhangInformationEfficientOffCenterSampling2021}. 
Sampling only at the most information
dense locations does, however, decrease the effective scanning area. MINFLUX provides a 
marked improvement in precision
over Gaussian-beam-based localization, but it has not been proven to provide the 
\textit{best} possible precision. Similarly, the Knight's Tour provides a better way of
achieving fast tracking than expanding the beam size and using the orbital method, but is
certainly not optimal in terms of precision. Gallatin and 
Berglund~\cite{gallatinOptimalLaserScan2012} answered the question of what the best scan
path is for optimal precision given a Gaussian beam of a certain size. In fact, all of the
work on this topic can be thought of as trying to answer the same question: What is the
optimal beam shape and scan pattern for a certain size scan area? In the case of MINFLUX, a
simpler version of this problem is considered where the size of the scan area is allowed
to be arbitrarily small. Gallatin and Berglund likewise simplified the question by assuming
a fixed beam shape~\cite{gallatinOptimalLaserScan2012}. Theirs remain the only conclusive answer 
to a version of the problem. An
alternative approach is not to have a fixed scan area at all but to optimize the scanning
pattern in real time. Vickers and Andersson~\cite{vickersInformationOptimalControl2021} 
showed that this can be done using a control law that optimizes the Fisher information in 
real time.

An important difference between the orbital method and the other methods is that the orbital
method makes use of continuous scanning. With the other methods, the beam has to move between 
spatial points on a significantly shorter time scale than the dwell time at each point (in our 
simulations, this scanning time was assumed to be instantaneous). This requires much faster 
scanning hardware than for the orbital method, thus adding significantly to the cost of 
implementation.

iSCAT has thus far only been applied once with real-time feedback, in the form of an ABEL 
trap~\cite{squiresInterferometricScatteringEnables2019}. In this work, we have shown that the 
method should outperform fluorescence-based localization for many samples, even ones that are 
intrinsically fluorescent. This finding makes it a very promising localization technique for 
RT-FD-SPT, which we hope will see much use in future. 

We have considered here only a subset of RT-FD-SPT methods, and only in two 
dimensions. The approach used can easily be applied to other methods such as 
other scanning patterns, multi-detector methods, or extended to three dimensions. It would also 
be very useful for investigating different control systems --- extremum seeking 
controllers~\cite{ashleyTrackingSingleFluorescent2016,ashleyControlLawSeeking2016}, for 
example --- and different estimation and filtering methods.  An important aspect of performance not addressed in this work is that of application-specific performance, notably with regard 
to spectroscopy. As we discuss in Ref.~\onlinecite{vanheerdenRealTimeFeedbackDrivenSingleParticle2021},
good performance for spectroscopy generally requires an adequate tracking precision to keep the
particle in the observation volume, together with a sufficient tracking duration to collect the necessary
spectroscopic data. Yet, it is conceivable that an analysis of different methods could be made based on the CRB of some measured spectroscopic quantity or even a simpler parameter such as the diffusion coefficient to find the truly optimal tracking method for a specific application.

To conclude, we have compared different RT-FD-SPT methods directly, and have found that there
is a fundamental trade-off between tracking speed and precision. The Knight's Tour method is the
fastest, MINFLUX is the most precise, and the orbital method is a cost-effective option
in between. iSCAT shows improved performance over fluorescence for
most of the biological examples considered in this study when using a suitably large illumination intensity.
The results are useful for choosing between methods and as a framework for comparing other RT-FD-SPT
methods.

\section*{Supplementary Material}

See supplementary material for calculation values of iSCAT parameters and details on the dynamical simulations.

\section*{Acknowledgements}

T.P.J.K. was supported by the South African Department of Science and Innovation's
National Research Foundation (NRF), Grant Nos. 87990, 94107, and 120387, and the Rental
Pool Programme of the National Laser Centre. B.v.H. was supported by the NRF Grant Nos.
115463 and 120387, and the South African Academy for Science and Art.

\section*{Data availability}
The data and code for all calculations and simulations is available at \url{https://github.com/bvanheerden/spt_sim}.

\bibliography{bibliography} 


\end{document}